\documentclass{aa}

\usepackage{graphicx}
\usepackage{txfonts}
\usepackage{epsfig}

\begin{document}
\title{Discovery of new Milky Way star cluster candidates in the
2MASS Point Source Catalog IV. Follow-up observations of cluster
candidates in the Galactic plane}
\subtitle{}

\author{V.\,D.~Ivanov\inst{1}
\and 
J.~Borissova\inst{1}
\and
F.~Bresolin\inst{2}
\and
P.~Pessev\inst{3}
}

\offprints{V.\,D.~Ivanov}

\institute{
    European Southern Observatory, Ave. Alonso de
    Cordova 3107, Casilla 19, Santiago 19001, Chile\\
    \email{vivanov, jborisso@eso.org}
  \and
    Institute for Astronomy, 2680 Woodlawn Drive, Honolulu, HI 96822,
    U.S.A.\\ 
    \email{bresolin@ifa.hawaii.edu}
  \and
    Department of Astronomy, Sofia University, Bulgaria.
    5~James Bourchier, 1164 Sofia, Bulgaria\\
    \email{pessev@phys.uni-sofia.bg}
  }

\date{Received .. ... 2004; accepted .. ... 2004}

\authorrunning{Ivanov et al.}
\titlerunning{Obscured Milky Way Star Clusters}

\abstract{
Nearly 500 cluster candidates have been reported 
by searches based on the new all-sky near infrared 
surveys. The true nature of the majority of these objects is 
still unknown. This project aims to estimate the physical 
parameters of some of the candidates in order to use them as 
probes of the obscured star formation in the Milky Way.

Here we report deep near infrared observations of four objects, 
discovered by our search based on the 2MASS Point Source Catalog
(Ivanov et al. \cite{iva02}; Borissova et al. \cite{bor03}).
CC\,04 appears to be a few million year old cluster. We estimate 
its distance and extinction, and set a limit on the total mass. 
CC\,08 contains red supergiants, indicating a slightly older age 
of about 7-10 Myr. The suspected cluster nature of CC\,13 was not 
confirmed. CC\,14 appears to be an interesting candidate with 
double-tail-like morphology but our data doesn't allow us to 
derive a firm conclusion about the nature of this object.

We found no supermassive star clusters similar to the Arches or 
the Quintuplet (M$_{tot}$$\geq$10$^4$\,M$_\odot$) among the 
dozen confirmed clusters studied so far in this series of papers, 
indicating that such objects are not common in the Milky Way.

}

\maketitle

\section{Introduction}

Young clusters are the sites of the most recent star formation 
in our Galaxy. They impact the galaxy evolution in numerous
ways, i.e. ionizing the surrounding gas, dissolving into the 
field population, and enriching the Galactic interstellar 
material. Clusters are excellent laboratories for star 
formation and stellar evolution research. However, they usually 
suffer from dust obscuration, rendering them invisible in the 
optical. The advent of infrared instrumentation that made 
it possible to carry out all-sky infrared surveys such as the 
Two Micron All Sky Survey (hereafter 2MASS, Skrutskie et al. 
\cite{scr97}) and the Deep Near Infrared Southern Sky Survey 
(DENIS, Epchtein et al. \cite{epc97}) offer an opportunity to 
investigate the most recent and up to now invisible star 
formation in the Milky Way. 

Simple visual inspection of IR images and automatic point 
source searches based on 2MASS and DENIS indicate that hidden 
clusters are surprisingly numerous, with nearly 500 new 
candidates. Yet, this valuable dataset remains relatively 
unutilized.
This paper is a part of a project that aims to determine the 
nature of these objects, and in particular, to answer the 
question of whether the Milky Way still forms massive clusters --
10$^4$ M$_\odot$ or higher, -- comparable to the Arches 
(Nagata et al. \cite{nag93}; see Figer et al. \cite{fig02} 
for latest review) and the Quintuplet clusters (Glass et al. 
\cite{gla87}; see also Figer et al. \cite{fig99}).

In Ivanov et al. (\cite{iva02}; hereafter Paper {\sc I}) we 
described our search algorithm and reported a list of cluster 
candidates discovered in the 2MASS Point Source Catalog. 
Borissova et al. (\cite{bor03}; hereafter Paper {\sc II}) 
reported additional candidates, and for the first time provided a 
study of the physical properties of one of our newly 
discovered cluster candidates CC\,01 (the caption of Fig. 1 
in that paper is wrong, the true sequence of the shown 
clusters is CC\,12, CC\,13, CC\,11, and CC\,14). 
In Borissova et al. (\cite{bor04}; hereafter Paper {\sc III}) 
we investigated the nature of clusters in a $\sim$10 degree 
region around the Galactic Center, selected from the catalogs
of Bica et al. (\cite{bic03a}) and Dutra et al. (\cite{dut03}).

This paper reports the results for four cluster candidates 
from our lists. They were selected because of their larger 
angular sizes and higher number of stars, in comparison with 
the other candidates (see the discussion in Paper {\sc II}). 
The possibility that they can populate the upper end of the 
cluster mass distribution motivated us to obtain estimates 
of their total mass.

\section{Observations and data reduction\label{SecObs}}

Near-infrared imaging follow-up of cluster candidates from 
Papers {\sc I} and {\sc II} were carried out on Dec 13, 2003
with the United Kingdom Infrared Telescope Fast Track Imager 
(UFTI; Roche et al. \cite{roc03}). The instrument uses a
1024\,$\times$\,1024 HgCdTe Hawaii array, with pixel scale 
of 0.091 arcsec pixel$^{-1}$. Table~\ref{TblObsLog} contains 
the observing log. The cluster candidates are shown in 
Figure~\ref{FigFields}. We refrained from creating a 
true-color image for CC\,08 because of the shallow $H$-band 
image.

\begin{table}[t]
\begin{center}
\caption{Log of the observations. The cluster candidates are 
identified by their numbers according to Papers {\sc I} 
and {\sc II}. For details see Sec.~\ref{SecObs}.}
\label{TblObsLog}
\begin{tabular}{lccc}
\hline
\multicolumn{1}{c}{ID} &
\multicolumn{1}{c}{R.A.~~~~~~~Dec.} &
\multicolumn{1}{c}{Filter}&
\multicolumn{1}{c}{Total Integration}\\
\multicolumn{1}{c}{CC} &
\multicolumn{1}{c}{(J2000.0)} &
\multicolumn{1}{c}{} & 
\multicolumn{1}{c}{Time, $\sec$}\\
\hline
\multicolumn{4}{c}{}\\
04 & 07:00:32~$-$08:52.0 & $J$        & 9\,$\times$\,2\,$\times$\,33\,=\, 594\\
   &                     & $H$        & 9\,$\times$\,4\,$\times$\,20\,=\, 720\\
   &                     & $K_S$      & 9\,$\times$\,6\,$\times$\,20\,=\,1080\\
\multicolumn{4}{c}{}\\
08 & 08:19:10~$-$35:39.0 & $J$        & 9\,$\times$\,2\,$\times$\,33\,=\, 594\\
   &                     & $H$        & 3\,$\times$\,4\,$\times$\,20\,=\, 240\\
   &                     & $K_S$      & 9\,$\times$\,6\,$\times$\,20\,=\,1080\\
\multicolumn{4}{c}{}\\
13 & 20:31:34~$+$45:05.8 & $K_S$      & 9\,$\times$\,6\,$\times$\,20\,=\,1080\\
\multicolumn{4}{c}{}\\
14 & 05:28:59~$+$34:23.2 & $J$        & 9\,$\times$\,2\,$\times$\,33\,=\, 594\\
   &                     & $H$        & 9\,$\times$\,4\,$\times$\,20\,=\, 720\\
   &                     & $K_S$      & 9\,$\times$\,6\,$\times$\,20\,=\,1080\\
\multicolumn{4}{c}{}\\
\hline
\end{tabular}
\end{center}
\end{table}

The data were taken in an alternating ``object''-``sky'' sequence,
starting and ending with a ``sky'' image. A total of nine ``object''
images and ten ``sky'' images were obtained, except for the the 
$H$-band of CC\,08 where the deteriorating weather conditions 
forced us to interrupt the sequence after only three ``object'' 
images were obtained. Each of these nine images is an average of 
2, 4, and 6 integrations of 33, 20, and 20 sec each, for 
$J$, $H$, and $K_S$, respectively. 

The first reduction step was to create a ``sky'' image by median 
combination of the sky images. We used two iterations, masking 
out the stars. It proved important because even the fields outside 
of our targets contain a significant number of stars. 
Dark-subtracted and normalized ``skies'' were used to flat-field the 
images. The flat fielding appeared sensitive to the position of 
the instrument, i.e. the pointing of the telescope. Therefore, we 
built separate ``sky'' and ``flat'' frames for each target and band. Next, 
the nine (three in the case of the $H$-band for CC\,08) individual 
``object'' images were aligned and combined.

\begin{figure}[t]
\caption{Near infrared images of \object{CC\,04}, \object{CC\,08}, 
\object{CC\,13}, and \object{CC\,14} (top left, top right, bottom 
left, and bottom right, respectively). For images with multiband 
observations the $J$, $H$, and $K_S$ are mapped onto blue, green 
and red, respectively. The individual images are 1.55 arcmin on the 
side. North is up, and East is to the left.}
\label{FigFields}
\end{figure}

The stellar photometry of these final
images was carried out using the ALLSTAR tool from DAOPHOT\,II 
(Stetson \cite{ste93}). We considered only stars with DAOPHOT 
errors smaller than 0.2 mag. This  mostly rejected faint stars,
affecting only our completeness limit. The median averaged internal 
photometric errors are $0.04\pm0.02$ for stars brighter than 17 
mag, and $0.08\pm0.04$ for the fainter ones. To account for the 
sky background variations we added to the individual errors in 
quadrature an additional $\sim$0.03 mag.

The presence of variable clouds made it impossible to calibrate 
the data with observations of photometric standards, and we used 
the 2MASS photometry of 10-25 stars for each pointing, selected 
to have no nearby companions in order to avoid crowding effects. 
The standard errors for the coefficients are typically 0.04-0.07 
for the zero points and 0.02-0.10 for the color terms. In summary, 
the conservative estimate of our total external photometric errors 
is 0.10-0.15 mag, not surprising given the crowded fields and the 
poor weather conditions.

\section{CC\,04\label{Sec_Obj04}}

This cluster candidate was selected based on the peak of the 
local stellar density in the 2MASS Point Source Catalog.
It is associated on the sky with a known reflection nebula (i.e. 
Racine \cite{rac68}). Houk \& Smith-Moore (\cite{hou88}) have 
determined the spectral type of a bright star in the vicinity - 
HD\,52329: A8\,{\sc V}. The morphology of 
the emitting gas suggests that this star is embedded in the
nebula. We used the known spectral type to determine the 
distance to CC\,04. Unfortunately HD\,52329 is at the edge 
of our field but the 2MASS Point Source Catalog lists 
$J$=8.865$\pm$0.021 and $K_S$=8.895$\pm$0.023 mag and
$J-K_S$=$-$0.03$\pm$0.03 mag. According to Bessell \& Brett 
(\cite{bes88}) $(J-K_S)_0$=$-$0.18 mag, so the E($J-K_S$)=0.11, 
$A_V$=0.65 and $A_K$=0.07 mag. Here, and throughout the rest 
of this paper we used the reddening law of Rieke \& Lebofsky 
(\cite{rie85}). Combining the absolute magnitude of 
$M_V\sim-4.4$ mag with $V-K_S$=$-$0.87 mag (Bessell \& Brett 
\cite{bes88}), we obtained $(m-M)_0$=12.6 mag. A similar 
exercise with the optical measurements of this star yields 
$(m-M)_0$=13.8 mag and $A_V$=0.52 mag. We adopted the average 
values of $(m-M)_0$=13.2 mag (d$\sim$4.4 Kpc) and $A_V$=0.6 mag.
Vogt (\cite{vog76}) used similar arguments to determine the 
distance to another star that appears associated with the 
cluster - ALS\,9249 (= LS\,VI\,$-$08\,2). He obtained 
$(m-M)_0$=13.5 mag (d$\sim$5 Kpc), close to our estimate. 
Racine (\cite{rac68}) gives a shorter distance of 
$(m-M)_0$=10.8 mag from the spectral type of BD$-$08\,1666 
but he only uses $UBV$ photometry to perform spectral 
classification, making his result less reliable than the
other ones.

\begin{figure}
\resizebox{\hsize}{!}{\includegraphics{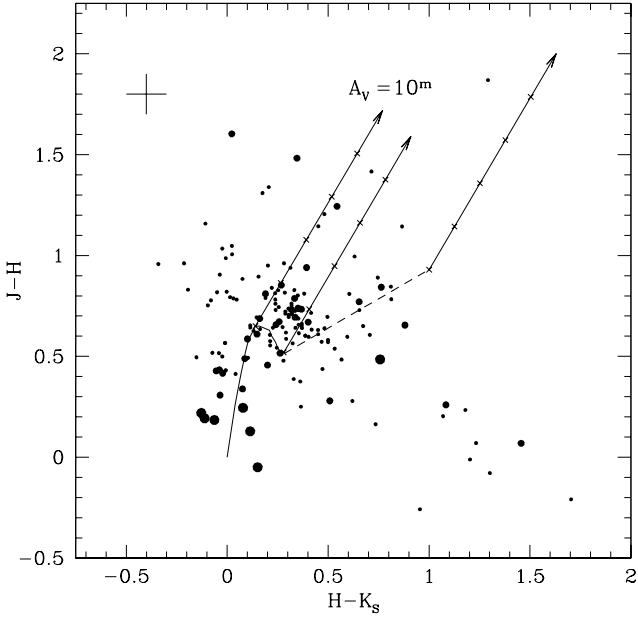}}
\caption{\object{CC\,04}. Color-color diagram $J-H$ versus 
$H-K_S$ from the UFTI data. The size of the dots is 
proportional to the apparent brightness of the stars. The 
cross in the upper left corner demonstrates the typical 
measurement errors.
The solid line shows the intrinsic colors of the main sequence 
stars from Frogel et al. (\cite{fro78}), extended to A0 
spectral class, assuming zero magnitude colors for the A0
star. 
The dashed line indicates the colors of unreddened T Tau stars 
from Meyer et al. (\cite{mey97}).
The reddening vectors for $A_V$\,=10 mag are also plotted. The 
interval between ticks correspond to $\Delta$$A_V$\,=2 mag.}
\label{Fig_CC04_CCDs}
\end{figure}

To verify the reddening determined above we used the 
color-color diagram (Figure~\ref{Fig_CC04_CCDs}). 
Assuming that the clump at $(H-K_S)\sim\,$0.3 mag, 
$(J-H)\sim\,$0.75 mag corresponds to the red end of the main 
sequence, we can estimate that the cluster suffers about 
0.5-1.5 mag of visual extinction. The size of the clump is 
comparable to the typical measurement uncertainty, and we 
adopted a tentative error in the reddening of 0.1 mag. 

\begin{figure}
\resizebox{\hsize}{!}{\includegraphics{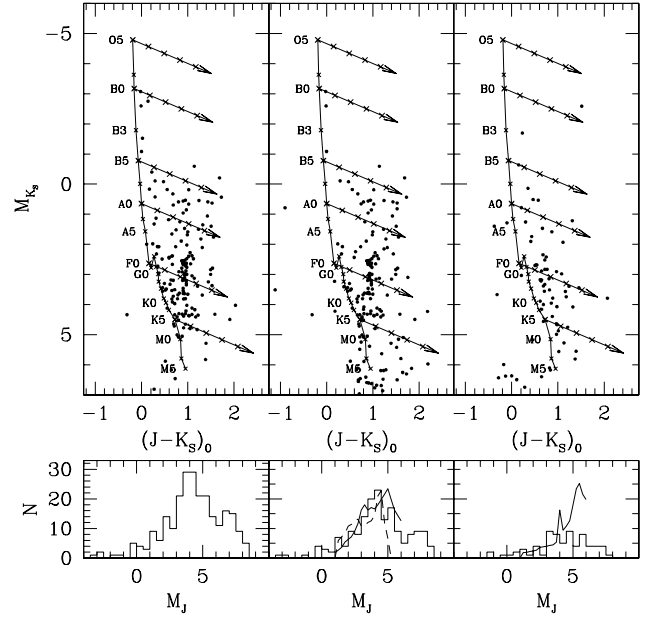}}
\caption{CC\,04: UFTI~ $M_{K_S}$ versus $(J-K_S)_0$ 
color-magnitude diagrams of all stars in the cluster field 
(top left), the stars remaining in the cluster field after 
the decontamination (top center), and all stars in the sky 
field (top right). 
The zero-age main sequence from Schmidt-Kaler (\cite{sch82}) 
is plotted, and the stellar spectral types are indicated on 
the left. 
The reddening vectors for $A_V$=10 mag are shown, with crosses 
spacing every 2 mag.
The corresponding $M_J$-band luminosity functions are also 
given (bottom row). The theoretical luminosity functions from 
Storm et al. (\cite{sto93}) for 1, 3, and 10 Myr are plotted 
(dashed and solid lines, bottom center; bottom right) for 
comparison. 
A distance modulus of $(m-M)_0$=13.2 mag and reddening of 
$A_V$=0.6 mag were adopted for all stars in both the cluster 
and the sky fields (see Sec.~\ref{Sec_Obj04} for details).
}
\label{Fig_CC04_NewCMD}
\end{figure}

The color-magnitude diagram $M_{K_S}$ versus $(J-K_S)_0$ 
of \object{CC\,04} is shown in Figure~\ref{Fig_CC04_NewCMD}. 
To determine the cluster mass we removed the fore- and 
back-ground stars statistically, as described in Paper 
{\sc III}. The result is also shown in the figure. Comparing
the decontaminated luminosity function with theoretical 
ones (Storm et al. \cite{sto93}) we derived an approximate 
cluster age 1-3 Myr. The small number statistics prevents us 
from more accurate determination. The field population 
appears older on average but this is only an approximation 
because we assigned to all field stars the distance and the 
reddening of the clusters, and in reality, they span a range 
of distances and reddening.

Finally, we determined the cluster mass following the same 
technique as in Paper {\sc III}. Unfortunately, the larger
cluster size in comparison with the UFTI field of view
prevented us from obtaining the total mass. Instead, we 
could only determine the mass of the part of the cluster 
that was covered by our image. We used the same technique 
as for CC\,01 (Paper {\sc II}) to derive the initial mass 
function (IMF) slope $\gamma$=$-$1.9$\pm$0.4, similar to 
the canonical Salpeter (\cite{sal55}) value of 
$\gamma$=$-$2.35 (i.e. Scalo \cite{sca86}, Eqs. 1.4, 1.7)
However, it should not be treated as evidence for the 
universality of the IMF because this method to derive the 
IMF has significant uncertainty. 

The sum of the masses of the stars with photometry is 
$\sim$200$\pm$70 solar masses. The error corresponds to a 
tentative uncertainty in the distance modulus of 0.5 mag. 
Integrating over the derived initial mass function down to 
0.8 solar mass stars we obtain a total cluster mass of 1200 
solar masses. As we explained in Paper {\sc III}, this is
only an upper limit because we overestimate the mass 
confined in stars with sub-solar mass. The uncertainty of 
this number easily reaches a factor of two, because of the 
errors in distance, reddening and the IMF slope. As we 
pointed out, some mass is ``missing'' due to the smaller 
UFTI field of view ($\sim$1.5$\times$1.5 arcmin, $\sim$2.25 
square arcmin area) in comparison with the bin-size of our 
2-dimensional histogram (2$\times$2 arcmin, 4 square arcmin 
area; see Paper {\sc I} for details). The 2MASS data indicate
that the stellar overdensity of CC\,04 is confined to one 
bin. Therefore, the total cluster mass is probably no 
larger than 2400 solar masses.

\section{CC\,08\label{Sec_Obj08}}

This object was discovered by our automatic algorithm. The 
2MASS color-magnitude diagram clearly indicated overdensity 
(Paper {\sc I}, Fig. 3). There is no associated radio, 
mid-infrared or extended nebular emission, indicating a
somewhat older age than the majority of the newly found 
cluster candidates. About half a dozen of the brightest 
potential members are visible on the DSS image. 

\begin{figure}
\resizebox{\hsize}{!}{\includegraphics{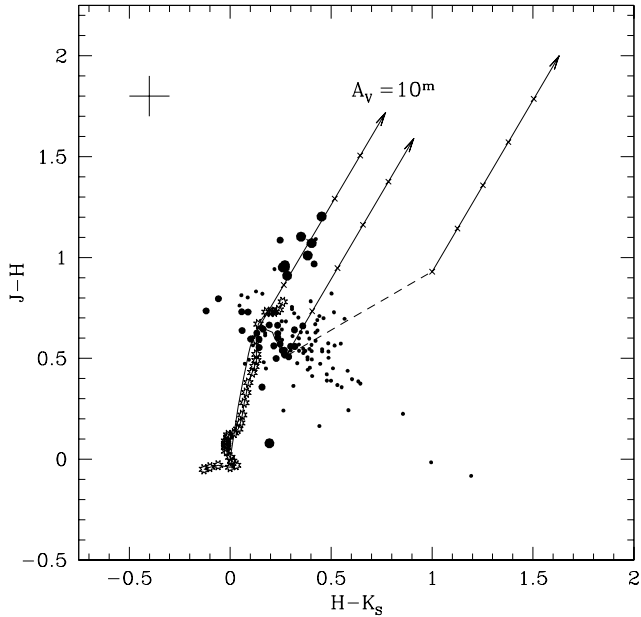}}
\caption{\object{CC\,08}. Color-color diagram $J-H$ versus 
$H-K_S$ from the UFTI data. 
The sequence of red supergiants from Bessell \& Brett
(\cite{bes88}) is shown with star symbols. 
See Figure~\ref{Fig_CC04_CCDs} for further explanations.}
\label{Fig_CC08_CCDs}
\end{figure}

The extinction estimate was obtained from the color-color
diagram (Figure~\ref{Fig_CC08_CCDs}). The presence of a 
clump of bright red objects is another notable difference. 
We interpret these objects as red supergiant stars (see for example 
Fig.\,7 in Vallenari et al. \cite{val00}). Their presence 
allows us to constrain the age of the cluster to 7-10 Myr, 
the evolutionary stage dominated by these stars, consistent 
with the lack of extended gas emission. CC\,08 appears to 
be an older object, perhaps a distant open cluster.

\begin{figure}
\resizebox{\hsize}{!}{\includegraphics{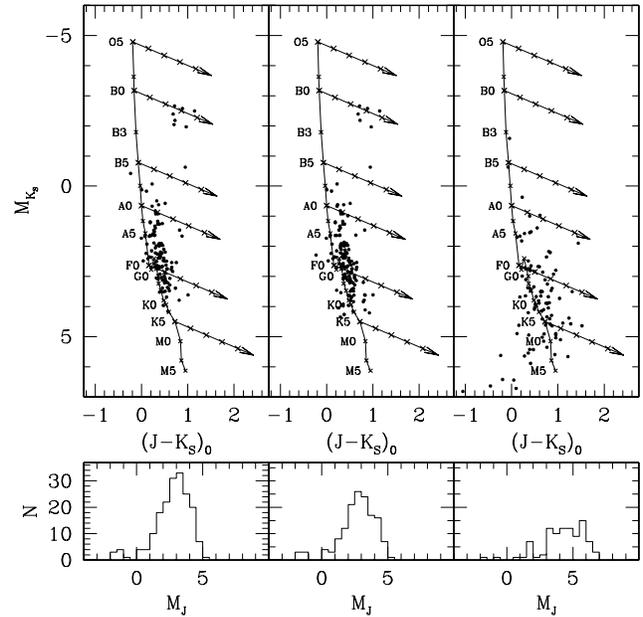}}
\caption{CC\,08: UFTI~ $M_{K_S}$ versus $(J-K_S)_0$ 
color-magnitude diagrams.
The details are identical to Figure~\ref{Fig_CC04_NewCMD}, 
except for the omitted theoretical luminosity functions.
A distance modulus of $(m-M)_0$=13.0 mag and reddening of 
$A_V$=3.0 mag were adopted for all stars in both the 
cluster and the sky fields (see Sec.~\ref{Sec_Obj08}).
}
\label{Fig_CC08_NewCMD}
\end{figure}

The photometry alone is not sufficient to determine the 
distance to the cluster. The reddening and the color 
magnitude diagram can only set a lower limit to the 
distance modulus of $(m-M)_0$$\sim$13.0 mag (d$\geq$4 Kpc)
and $A_V$=3.0 mag using the unreddened zero-age main 
sequence as a blue limit of the cluster members 
(Figure~\ref{Fig_CC08_NewCMD}). The distance modulus can 
easily be 2 mag higher. For completeness we performed the 
decontamination but the lack of a distance estimate 
prevented us from further analysis. 

\section{CC\,13}

{\object{CC\,13} was also selected during our overdensity 
search. It coincides with an H{\sc ii} region. The object was 
imaged only in $K_S$. The deeper photometry did not confirm our 
original result based on the 2MASS data. The luminosity 
function (Figure~\ref{Fig_CC13_LF}) of the stars in the deep 
UFTI image appears to show no significant excess over the field. 
The difference in the bright end of the two luminosity 
functions -- the excess of the bright stars in the field of the 
candidate -- explains why it was selected: CC\,13 appears to be 
a concentration of about half a dozen hot stars that ionize the 
surrounding gas, similar to the well-known OB associations in 
the Milky Way and nearby galaxies (i.e. Bresolin et al. 
\cite{bre98}, Ivanov et al. \cite{iva96}). We are reluctant to 
classify such a poor group as a cluster.

\begin{figure}
\resizebox{\hsize}{!}{\includegraphics{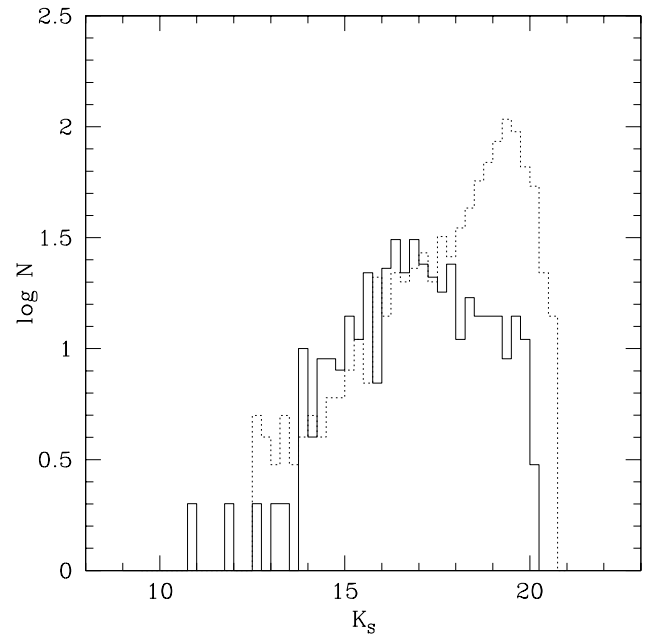}}
\caption{\object{CC\,13}. Luminosity functions of the field 
of the cluster candidate (solid line) and a nearby sky field 
(dotted line). 
}
\label{Fig_CC13_LF}
\end{figure}

\section{CC\,14}

\object{CC\,14} was discovered serendipitously during a visual 
inspection of the region around another cluster candidate as a 
group of anomalously red stars, with two East-West tails. The 
new $K$-band image indicates a cluster diameter of about 1 
arcmin. None of the suspected member stars is visible on the 
DSS suggesting heavy extinction. The H{\sc II} region 
[KC97c]\,G173.5$-$00.1 is located $\sim$1.5 arcmin away from 
\object{CC\,14}. Fich \& Blitz (\cite{fic84}) used the radial 
velocity and a Galactic rotation model to place it at a 
distance of 2.3$\pm$0.7 Kpc or $(m-M)_0$=11.8$\pm$0.7 mag. 
Note that there is no evidence for a direct physical connection 
between the H{\sc II} region and the cluster candidate.

\begin{figure}
\resizebox{\hsize}{!}{\includegraphics{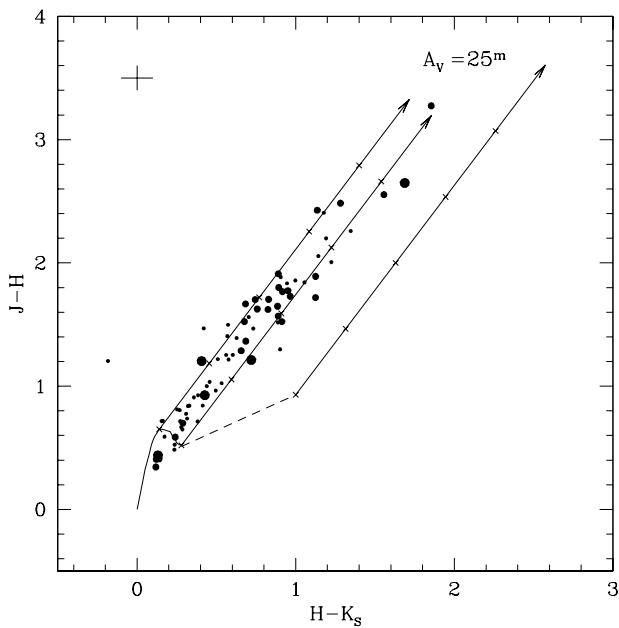}}
\caption{\object{CC\,14}. Color-color diagram $J-H$ versus 
$H-K_S$ from the UFTI data. 
See Figure~\ref{Fig_CC04_CCDs} for further explanations.
The ticks on the reddening vectors here span 5 mag 
intervals.}
\label{Fig_CC14_CCDs}
\end{figure}

The color-color diagram (Figure~\ref{Fig_CC14_CCDs}) indicates 
that the cluster - marked by the clump of brighter stars at 
$H-K$$\sim$0.9 and $J-H$$\sim$1.7 mag, - is subjected to 13-15 
mag of optical extinction. To determine the physical parameters 
of the cluster we adopted the method of the 10th bright star, 
proposed by Dutra \& Bica (\cite{dut01}), and based on absolute 
magnitudes and colors of hot main-sequence stars from Cotera 
et al. (\cite{cot00}). The assumption that CC\,14 is a massive 
cluster, similar to NGC\,3603, means that the 10th brightest 
cluster member is an O5\,V, yielding $(m-M)_0$=17.3$\pm$0.4, 
$A_V$=15.0$\pm$0.3, and $A_K$=1.68$\pm$0.03 mag. Here the 
uncertainties are dominated by the number statistics, and they 
are estimated by comparing the parameters derived from the 10th 
and from the 13th star, as explained in Paper {\sc III}. If 
CC\,14 is less massive, and the 10th star is a B0\,V, then 
$(m-M)_0$=15.8$\pm$0.4, $A_V$=15.1$\pm$0.3, and 
$A_K$=1.69$\pm$0.03 mag. We adopted tentatively the average of 
the two: $(m-M)_0$=16.6$\pm$0.8 mag, and $A_V$=15.1$\pm$0.3 
mag. We stress once again that these are only approximate 
estimates. The color-magnitude diagram of CC\,14 is shown in 
Figure~\ref{Fig_CC14_NewCMD}. We removed the background using 
the sky frame. Note that the field stars seems to  better fit
the main sequence for $A_V$=8.0 mag, which may correspond to 
the foreground extinction.

\begin{figure}
\resizebox{\hsize}{!}{\includegraphics{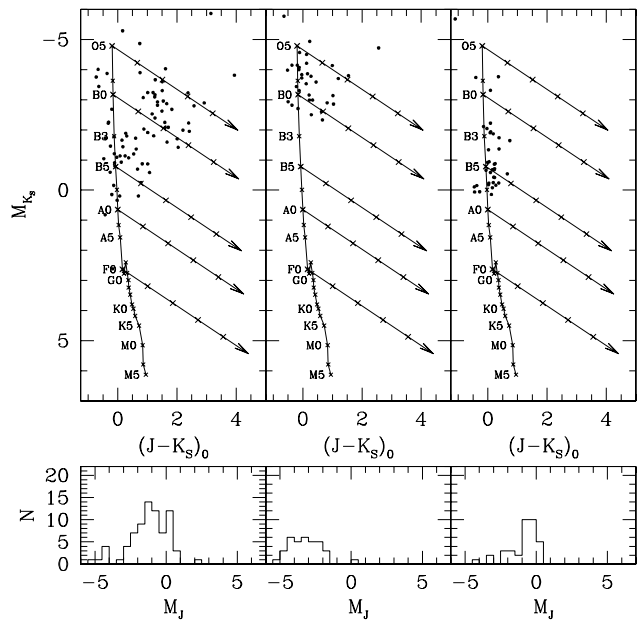}}
\caption{CC\,14: UFTI~ $M_{K_S}$ versus $(J-K_S)_0$ 
color-magnitude diagrams.
The details are identical to Figure~\ref{Fig_CC04_NewCMD}, 
except for the omitted theoretical luminosity functions.
Distance modulus of $(m-M)_0$=15.8 mag and reddening of 
$A_V$=15.0 mag were adopted for the stars on the 
decontaminated diagram. The stars in the other two panels
were de-reddened with $A_V$=8.0 mag (see Sec.~\ref{Sec_Obj08}).
}
\label{Fig_CC14_NewCMD}
\end{figure}

Imaging of a larger field -- encompassing the ``tidal'' tails -- 
and spectroscopic determinations of the spectral type for at 
least a few members are necessary to estimate the rest of the 
cluster parameters. The available data allow us only to 
determine that the mass of the detected stars is 1100-1300 
solar masses, assuming an age of 1 Myr. This estimate is hampered 
by severe uncertainties, including the limited field of view. 
CC\,14 could be a massive cluster, and it needs further 
investigation. The two tails (Paper {\sc I}) hint that we may 
be seeing a cluster in the middle of a disruption event or a 
case of sequential star formation where one of the groups has 
triggered the formation of the others via shock waves.

\section{Summary and conclusions}

We report deep $JHK_S$ imaging of four cluster candidates 
selected from our search (Papers {\sc I} and {\sc II}) 
based on the surface density in the 2MASS Point Source 
Catalog. Three of them appear to be young clusters. In one 
case we set a moderate upper mass limit. Another object is 
just a poor group of stars, and the third candidate is a 
more likely to be an open cluster. The total mass of the 
last object -- which appears cluster-like -- remains 
unknown. Our mass estimates are based on photometry alone 
which leaves the IMF slopes extremely ill-defined, with an 
uncertainty in the total mass up to a factor of two or 
three. More accurate mass measurements require spectroscopy. 
There are no obvious supermassive 
(M$_{tot}$$\geq$10$^4$\,M$_\odot$) objects among the dozen 
confirmed clusters that we have studies in this series of 
papers, suggesting that the Milky Way may lack a numerous 
hidden population of supermassive clusters.

The object CC\,13 is a conglomerate of hot stars, embedded 
in emitting gas but without an underlying overdensity of 
faint low-mass stars. The comparison with the images of the 
other objects excludes the possibility that the faint 
population is lost due to crowding. This result points out 
the necessity to formulate a robust definition of a cluster.
Porras et al. (\cite{por03}) postulates that a ``cluster''
has to contain at least five associated stars but this 
number has no physical justification, and it borders the 
number of components in some multiple stars. In the 
framework of the commonly used classification of stellar 
systems, CC\,13 is closer to sparse OB associations than to 
typical young clusters. We hesitate to classify similar 
objects as clusters, although we realize this is subject to 
further discussion. CC\,13 and similar objects may present 
an interesting example of star-forming sites where the 
appearance of the first massive hot stars has truncated the 
formation of low-mass stars. Deeper observations are 
necessary to derive a firmer conclusion about the true 
nature of this object. 

Finally, we underline that spectroscopy can  greatly improve
the cluster mass, distance and reddening estimates, in 
comparison with the current ones, based only on photometry.

\begin{acknowledgements}
This publication makes use of data products from the 2MASS, 
which is a joint project of the University of Massachusetts 
and the IPAC/CalTech, funded by the NASA and the NSF. This 
research has made use of the SIMBAD database, operated at CDS, 
Strasbourg, France. We are grateful to the UKIRT staff for 
their assistance. UKIRT is operated by the JAC on behalf of 
the U.K. PPARC. We thank the anonymous referee for the 
suggestions that helped to improve the paper.
\end{acknowledgements}


\begin{thebibliography}{}
\bibitem[1988]{bes88} Bessell, M.S. \& Brett, J.M. 1988, \pasp, 
        100, 1134
\bibitem[2003]{bic03a} Bica, E., Dutra, C.M. \& Barbuy, B. 2003, 
        \aap, 397, 177
\bibitem[2003]{bor03} Borissova, J., Pessev, P., Ivanov, V.D., 
        Saviane, I., Kurtev, R., Ivanov, G.R. 2003, \aap, 411, 83 
        (Paper {\sc II})
\bibitem[2004]{bor04} Borissova, J., Ivanov, V.D., Minitti, D., 
        Geisler, D., \& Stephens, A. 2004, \aap, accepted
        (Paper {\sc III})
\bibitem[1998]{bre98} Bresolin, F., Kennicutt, R.C., Jr., Ferrarese, L., 
        Gibson, B.K., Graham, J.A., Macri, L.M., Phelps, R.L., Rawson, 
        D.M., Sakai, S., Silbermann, N.A., Stetson, P.B., Turner, A.M.
        1998, \aj, 116, 119
\bibitem[2000]{cot00} Cotera, A.S., Simpson, J.P., Erikson, E.F., 
        et al. 2000, \apjs, 129, 123
\bibitem[2001]{dut01} Dutra, C.M. \& Bica, E. 2001, \aap, 376, 434
\bibitem[2003]{dut03} Dutra, C.M., Bica, E., Soares, J. \& Barbuy, B. 
        2003, \aap, 400, 533
\bibitem[1997]{epc97} Epchtein, N. 1997, in ASSL Vol. 210, The Impact 
        of Large Scale Near-IR Sky Surveys, ed. F. Garzon et al. 
        (Dordrecht: Kluwer), 15
\bibitem[1984]{fic84} Fich, M., \& Blitz, L. 1984, \apj, 279, 125
\bibitem[1999]{fig99} Figer D.F., McLean, I.S., \& Morris, M. 1999, 
        \apj, 514, 202
\bibitem[2002]{fig02} Figer D.F., Najarro, F., Gilmore, D., Morris, M.,
        Kim, S.S., et al. 2002, \apj, 581, 258
\bibitem[1978]{fro78} Frogel, J.A., Persson, S.E., Matthews, K. \& 
        Aaronson, M. 1978, \apj, 220, 75
\bibitem[1987]{gla87} Glass, I.S., Catchpole, R.M., \& Whitelock,
        P.A. 1987, \mnras, 227, 373
\bibitem[1988]{hou88} Houk, N. \& Smith-Moore, M. 1988, Michigan 
        Spectral Survey, Ann Arbor, Dept. of Astronomy, Univ. 
        Michigan, Vol. 4
\bibitem[1996]{iva96} Ivanov, G.R. 1996, \aap, 305, 708
\bibitem[2002]{iva02} Ivanov, V.D., Borissova, J., Pessev, P., 
        Ivanov, G.R., Kurtev, R. 2002, \aap, 394, 1 (Paper {\sc I})
\bibitem[1997]{mey97} Meyer, M., Calvet, N., Hillenbrand, L., 1997, 
        \aj, 114, 228
\bibitem[1993]{nag93} Nagata, T., Hyland, A.R., Straw, S.M., Sato, S., 
        \& Kawara, K. 1993, \apj, 406, 501
\bibitem[2003]{por03} Porras, A., Christopher, M., Allen, L., 
        DiFrancesco, J, Megeath, S. Th., Myers, P. 2003, \aj, 126, 
        1916
\bibitem[1968]{rac68} Racine, R. 1968, \aj, 73, 588
\bibitem[1985]{rie85} Rieke, G.H. \& Lebofsky, M.J. 1985, \apj, 288, 
        618
\bibitem[2003]{roc03} Roche, P.F., Lucas, P.W., Mackay, C.D., 
        Ettedgui-Atad, E., Hastings, P.R., Bridger, A., Rees, N.P., 
        Leggett, S.K., Davis, C., Holmes, A.R., Handford, T. 2003, 
        SPIE, 4841, 901
\bibitem[1955]{sal55} Salpeter, E.E. 1955, \apj, 121, 161
\bibitem[1986]{sca86} Scalo, J.M. 1986, Fund. of Cosm. Phys., 11, 1
\bibitem[1982]{sch82} Schmidt-Kaler, T., 1982, in Landolt-Borstein,
        New Series, Group VI, vol. 2, ed. K. Schaifers \& H.H. Voigt
        (Berlin: Springer-Verlag),1
\bibitem[1997]{scr97} Skrutskie, M.F., et al. 1997, in ASSL Vol. 210, 
        The Impact of Large Scale Near-IR Sky Surveys, ed. F. Garzon 
        et al. (Dordrecht: Kluwer), 25
\bibitem[1993]{ste93} Stetson, P. B. 1993, User's Manual for 
        {\sc daophot ii}
\bibitem[1993]{sto93} Storm, K., Strom, S., Merrill, M., 1993, \apj, 233
\bibitem[2000]{val00} Vallenari, A., Carraro, G., \& Richichi, A. 
        2000, \aap, 353, 147
\bibitem[1976]{vog76} Vogt, N. 1976, \aap, 53, 9
\end{thebibliography}
\end{document}